\documentclass[lettersize,journal]{IEEEtran}
\usepackage{amsmath,amsfonts}
\usepackage{algorithm}
\usepackage{array}
\usepackage{stfloats}
\usepackage{url}
\usepackage{verbatim}
\usepackage{graphicx}
\usepackage{cite}
\usepackage{multirow}
\usepackage{color}

\usepackage{newtxtext,newtxmath}
\usepackage[noend]{algpseudocode}
\algnewcommand{\Inputs}[1]{%
  \State \textbf{Inputs:}~\parbox[t]{.8\linewidth}{\raggedright #1}
}
\algnewcommand{\Initialize}[1]{%
  \State \textbf{Initialize:}
}
\algnewcommand{\Def}[1]{%
  \State \textbf{Procedure}~\parbox[t]{.8\linewidth}{\raggedright #1}
}
\newcommand{\ind}{\hspace*{\algorithmicindent}}

\makeatletter
\def\caption@documentclass{elsarticle}%
\makeatother

\usepackage[hang,small,bf]{caption}
\usepackage[subrefformat=parens]{subcaption}
\begin{document}

% \title{Efficient Speech Segmentation with Pre-trained\\ Speech Encoder for Speech Translation}
% \title{Improving Accuracy and Time Efficiency of Speech Translation by Speech Segmentation using Fine-tuned wav2vec 2.0}
\title{Improving Speech Translation Accuracy and Time Efficiency with Fine-tuned wav2vec 2.0-based Speech Segmentation}
% Improved speech translation accuracy and efficiency with fine-tuned wav2vec 2.0-based speech segmentation model
% self-supervised speech models

\author{Ryo Fukuda,~\IEEEmembership{Nonmember},~Katsuhito Sudoh,~\IEEEmembership{Nonmember},~Satoshi Nakamura,~\IEEEmembership{Fellow, IEEE}
        % <-this % stops a space
\thanks{Ryo Fukuda is with the Graduate School of Science and Technology, Nara Institute of Science and Technology, Ikoma 630-0192, Japan (e-mail: fukuda.ryo.fo3@is.naist.jp).}
\thanks{Katsuhito Sudoh is with the Graduate School of Science and Technology, Nara Institute of Science and Technology, Ikoma 630-0192, Japan (e-mail: sudoh@is.naist.jp).}
\thanks{Satoshi Nakamura is with the Data Science Center and Graduate School of Science and Technology, Nara Institute of Science and Technology, Ikoma 630-0192, Japan (e-mail: s-nakamura@is.naist.jp).}
}

% \markboth{IEEE/ACM Transactions on Audio, Speech, and Language Processing}


\maketitle

\begin{abstract}
% Speech segmentation, which splits long speech into short segments, is essential for speech translation (ST).
Speech translation (ST) automatically converts utterances in a source language into text in another language.
Splitting continuous speech into shorter segments, known as speech segmentation, plays an important role in ST.
Recent segmentation methods trained to mimic the segmentation of ST corpora have surpassed traditional approaches.
Tsiamas et al.~\cite{tsiamas22_interspeech} proposed a segmentation frame classifier (SFC) based on a pre-trained speech encoder called wav2vec 2.0.
Their method, named SHAS, retains 95-98\% of the BLEU score for ST corpus segmentation.
However, the segments generated by SHAS are very different from ST corpus segmentation and tend to be longer with multiple combined utterances.
This is due to SHAS's reliance on length heuristics, i.e., it splits speech into segments of easily translatable length without fully considering the potential for ST improvement by splitting them into even shorter segments.
Longer segments often degrade translation quality and ST's time efficiency.
In this study, we extended SHAS to improve ST translation accuracy and efficiency by splitting speech into shorter segments that correspond to sentences.
We introduced a simple segmentation avlgorithm using the moving average of SFC predictions without relying on length heuristics and explored wav2vec 2.0 fine-tuning for improved speech segmentation prediction.
Our experimental results reveal that our speech segmentation method significantly improved the quality and the time efficiency of speech translation compared to SHAS.
\end{abstract}


\section{Introduction}
\label{sec:introduction}
\IEEEPARstart{T}{he} segmentation of continuous speech is a fundamental process required for speech translation (ST) and other spoken language applications.
In text-to-text machine translation (MT), the input text is usually segmented into sentences using punctuation marks as boundaries.
However, such explicit boundaries are unavailable in ST.
ST corpora usually contain speech segments that are aligned to sentences.
For example, the procedure for creating a multilingual ST corpus, MuST-C \cite{di-gangi-etal-2019-must}, first performs sentence alignment between English transcriptions and its translations and aligns the English speech and transcriptions with a forced aligner.
Much ST research uses such sentence-aligned speech segments to train and evaluate systems, although they cannot be used in practical situations.
In addition, existing ST models cannot directly translate long continuous speech without segmentation.
One reason is that the required computational resources increase with the length of the input speech.
Even without any constraints on computational resources, an ST model trained on segmented short speech struggles to translate extremely long speech that is not included in its training data.
For these reasons, several efforts have focused on speech segmentation for ST.
\par
\textbf{Pause-based segmentation} with voice activity detection (VAD) is commonly used as preprocessing for automatic speech recognition (ASR) and ST.
However, pauses in speech do not necessarily coincide with boundaries of semantic units such as sentences in text, e.g., there may be long pauses in an utterance corresponding to a sentence or almost no pauses between utterances.
Over-segmentation, in which a silence interval fragments a sentence, and under-segmentation, in which multiple sentences are included in one segment while ignoring a short pause, reduce the ASR and ST performances \cite{wan2021segmenting}.
Fixed-length segmentation is the simplest approach that segments audio at a predefined segment length \cite{sinclair2014semi}.
There is also a combination method that concatenates speech segments generated by VAD up to a certain length.
Such \textbf{length-based segmentation} methods are heuristic approaches that can split speech into segments of easily translatable length \cite{gaido-etal-2021-beyond}.
\textbf{Punctuation-based segmentation} methods are often used in cascade ST, re-segmenting the ASR results of segments produced by VAD with a punctuation restoration model or a language model \cite{4518807,cho-etal-2012-segmentation}.
These methods can improve the translation accuracy of MT, but they cannot be used for end-to-end ST, where the source language is translated directly without ASR.
In addition, these methods cannot prevent ASR errors due to improper segmentation.
\par
As mentioned above, ST corpora usually have speech segments that correspond to sentences, which are suitable for translation.
Recent \textbf{corpus-based segmentation} methods have been successful using a classification model trained to predict segmentation of ST corpora.
A corpus-based method, SHAS~\cite{tsiamas22_interspeech}, led to state-of-the-art results with a segmentation frame classifier (SFC) based on a pre-trained speech encoder called wav2vec 2.0~\cite{baevski2020wav2vec}.
However, the segments generated by SHAS tend to be significantly longer than segments of ST corpus.
Such long segments can decrease translation quality.
In addition, the longer the segment is, the more computation time required for translation.
These long segments are caused by using segmentation algorithms that place more importance on the lengths of segments than SFC prediction.
This strategy allows SHAS to split speech into segments whose lengths are preferred by ST.
However, the following potential remains unconsidered: improving ST translation accuracy and time efficiency by splitting these segments even shorter.
\par
In this work, we extend SHAS to improve ST translation accuracy and efficiency by splitting speech into shorter segments that correspond to sentences, such as those included in ST corpus.
We introduce a simple segmentation algorithm using the moving average of SFC predictions without relying on length heuristics to produce shorter segments.
We also introduce an efficient fine-tuning of wav2vec 2.0 to improve SFC accuracy.
We conducted experiments with an end-to-end ST on MuST-C v2 for English-to-German.
Our experimental results showed that the proposed method retained 97.4\% of BLEU score for MuST-C segmentation included in the corpus, surpassing 95.1\% by SHAS.
We also showed that the proposed method reduced the translation time by about 20\% while improving translation accuracy by generating shorter segments.
Our case analysis revealed that our proposed method sometimes outperformed MuST-C segmentation and produced competitive translation results.
Furthermore, an evaluation using 8 language pairs from MuST-C v1 and Europarl-ST showed that the proposed method is effective for target languages and domains that differ from the dataset used to train the SFC.

\section{Related work}
Early studies on segmentation for ST considered modeling with the Markov decision process \cite{mansour2010morphtagger, sinclair2014semi}, conditional random fields \cite{nguyen-vogel-2008-context, lu2010better}, and support vector machines \cite{diab-etal-2004-automatic, sadat-habash-2006-combination, matusov-etal-2007, rangarajan-sridhar-etal-2013-segmentation}.
They focused on cascade ST systems that consist of an ASR model and a statistical machine translation  model, which were superseded by newer ST systems based on neural machine translation.
\par
In recent studies, many speech segmentation methods based on VAD have been proposed for ST.
Gaido et al. \cite{DBLP:journals/corr/abs-2104-11710} and Inaguma et al. \cite{inaguma-etal-2021-espnet} used the heuristic concatenation of VAD segments up to a fixed length to address the over-segmentation problem.
Gállego et al. \cite{gallego2021end} used a pre-trained ASR model called wav2vec 2.0 \cite{baevski2020wav2vec} for silence detection.
Yoshimura et al. \cite{yoshimura2020end} used an RNN-based ASR model to consider consecutive blank symbols (``\_") as a segment boundary in decoding using connectionist temporal classification (CTC).
Such CTC-based speech segmentation has the following advantage; segment lengths can be intuitively controlled by adjusting the number of consecutive blank symbols that are regarded as segment boundaries.
However, these methods often split speech at inappropriate boundaries for ST because they mainly segment speech based on long pauses.
\par
Re-segmentation using ASR transcripts is widely used in cascade STs.
Improvements in MT performance have been reported by re-segmenting transcriptions to sentence units using punctuation restoration \cite{lu2010better, rangarajan-sridhar-etal-2013-segmentation, cho2015punctuation, ha2015kit, cho17_interspeech} and language models \cite{stolcke1996automatic, wang2016efficient}.
Unfortunately, they are difficult to use in end-to-end ST and cannot prevent the ASR errors caused by speech segmentation.
\par
Corpus-based segmentation using manually or semi-manually segmented speech corpora is a leading recent approach.
Wan et al. \cite{wan2021segmenting} introduced a re-segmentation model trained with movie and TV subtitle corpora to modify the segment boundaries in ASR output.
Wang et al. \cite{wang2019online} and Iranzo-Sánchez et al. \cite{iranzo-sanchez-etal-2020-direct} proposed an RNN-based text segmentation model trained with a bilingual speech corpus.
Methods for directly segmenting speech with a segmentation model have also recently been proposed \cite{fukuda22b_interspeech, tsiamas22_interspeech}.
Fukuda et al.~\cite{fukuda22b_interspeech} used a Transformer encoder to build a segmentation model and also proposed a hybrid method that combines VAD and the prediction of the segmentation model.
Tsiamas et al.~\cite{tsiamas22_interspeech} built an SFC based on a pre-trained speech encoder called wav2vec 2.0.
Their method, Supervised Hybrid Audio Segmentation (SHAS), is the current state-of-the-art method of speech segmentation for ST.

In our approach, we improve the accuracy of the SFC by unfreezing parts or all of the wav2vec 2.0 parameters during training. Performing full fine-tuning can significantly increase the training cost compared to the original SHAS, so we use a method of parameter-efficient transfer learning (PETL). PETL is a research direction aimed at reducing the computational costs of applying large pre-trained models to new tasks \cite{houlsby2019parameter,he2021towards,sung2022lst}.

\section{Review of SHAS}
\label{sec:shas}
In this section, we describe an SFC (\ref{subsec:sfc}) and a probabilistic divide-and-conquer (pDAC) algorithm (\ref{subsec:pdac}) of the state-of-the-art speech segmentation method called SHAS.
Then we describe its drawback: producing lengthy segments (\ref{subsec:problem}).

\subsection{Segmentation frame classifier}
\label{subsec:sfc}

\begin{figure}[t]
\centering
\includegraphics[keepaspectratio, width=7cm]{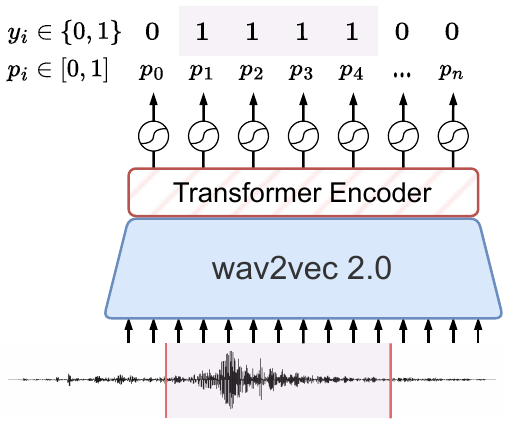}
\caption{SFC of SHAS: Value of $y=1$ indicates that corresponding frame is part of a segment of ST corpus, and $y=0$ indicates that it is part of a segment boundary.}
\label{fig:sfc}
\end{figure}

The SFC determines whether each input speech frame belongs to a segment or a segment boundary.
It is implemented as a neural network model with a single Transformer encoder layer that is connected to the encoder of the pre-trained, self-supervised speech model, wav2vec 2.0 (Fig. \ref{fig:sfc}).
Given an ST corpus, each frame of speech is labeled as 1 or 0, depending on whether it is included in a segment.
During training, speech segments of N seconds split at random positions are used with sequences of labels $y \in \{0,1\}$ that correspond to model output sequences.
$y=1$ indicates that the corresponding frame is inside a segment, and $y=0$ indicates that it is outside of the segments, i.e., belonging to a segment boundary.
% The parameters of the wav2vec 2.0 are kept fixed during the training, with only the parameters of the final Transformer Encoder layer and the output layer being updated.

% At the inference time, given an unlabeled audio waveform, it is split into length N and input to the SFC without overlapping.
At the inference time, given an unlabeled audio waveform, it is split into contiguous segments of length N, which are then input to the SFC. These segments are arranged in such a way that there is no temporal overlap between consecutive segments.
For each input with length N, SFC predicts probabilities corresponding to audio frames of length $n=N/320$ due to the convolutional feature extractor of wav2vec 2.0.
The wav2vec 2.0 parameters are kept fixed during the training, and only the parameters of the final Transformer encoder layer and the output layer are updated.

\subsection{Probabilistic divde-and-conquer}
\label{subsec:pdac}

\begin{algorithm}[ht]
\caption{pDAC}
\label{algo:pdac}
\begin{algorithmic}[1]
\Inputs{$probs,~max,~min,~thr$}
\Initialize\\
\State{\ind$segments \gets \text{empty List}$}
\State{\ind$sgm \gets \text{Tuple(0, $probs$.length)}$}
\Comment{Init single segment}
% \State{\ind$thrs[:min] \gets 0,~thrs[min:] \gets thr$}\\
% \Comment{Set threshold filter $thrs$}
% \State{\ind$probs \gets MovingAverage(probs,~n\_ma) $}\\
\State{RECURSIVE\_SPLIT($sgm$)}
\State{return $segments$}
\\
% \State{\hspace*{\algorithmicindent}\hspace*{\algorithmicindent}}
\Def{RECURSIVE\_SPLIT($sgm$)}
\State{\ind\textbf{if}~$sgm\text{.length} < max$ \textbf{then}}
\State{\ind\ind\textbf{append} $sgm$ to segments}
\State{\ind\textbf{else}}
\State{\ind\ind$j \gets 0$}
\State{\ind\ind$indices \gets \textbf{argsort}~ probs[sgm]$}
\State{\ind\ind\textbf{while} True \textbf{do}}
\State{\ind\ind\ind$sgm_a,~sgm_b \gets \textbf{split}~sgm~\textbf{at}~indices[j]$}
\State{\ind\ind\ind$sgm_a \gets \textbf{trim}(probs[sgm_a],~thr)$}
\State{\ind\ind\ind$sgm_b \gets \textbf{trim}(probs[sgm_b],~thr)$}
\State{\ind\ind\ind\ind$\textbf{if}~sgm_a\text{.length} > min~\textbf{and}$}
\State{\ind\ind\ind\ind$~~~sgm_b\text{.length} > min~\textbf{then}$}
%\State{\ind\ind\ind\ind\ind\ind$\textbf{and}~sgm_b\text{.length} > min~\text{then}$}
\State{\ind\ind\ind\ind\ind RECURSIVE\_SPLIT($sgm_a$)}
\State{\ind\ind\ind\ind\ind RECURSIVE\_SPLIT($sgm_b$)}
\State{\ind\ind\ind\ind\ind \textbf{break}}
\State{\ind\ind\ind\ind$j \gets j+1$}
\end{algorithmic}
\end{algorithm}

% \red{Based} on the probability of each frame being included in a segment as predicted by SFC, pDAC segments the audio.
During inference, pDAC divides speech based on the probability of each frame being included in a segment ($probs$) predicted by SFC.
pDAC is a recursive algorithm that splits speech at the point least likely to be in a segment and applies the same split to the two resulting segments (Algorithm \ref{algo:pdac}).
The algorithm utilizes three hyperparameters: a maximum segment length ($max$) to regulate the length of the resulting segments, a minimum segment length ($min$) to prevent excessively small noisy segments, and a threshold ($thr$) to trim a segment's ends, which are classified as being excluded from segments.
After a split, the resulting segments are trimmed to the first and last frames $i, j$ with $p_i, p_j > thr$.
A split is performed until the segment's length is less than $max$.
This allows pDAC to keep the segments’ length within a certain range.

\subsection{Lengthy segments by SHAS}
\label{subsec:problem}

\begin{figure}[!t]
\centering
\subfloat[Sentence-aligned speech segmentation]{\includegraphics[width=9cm]{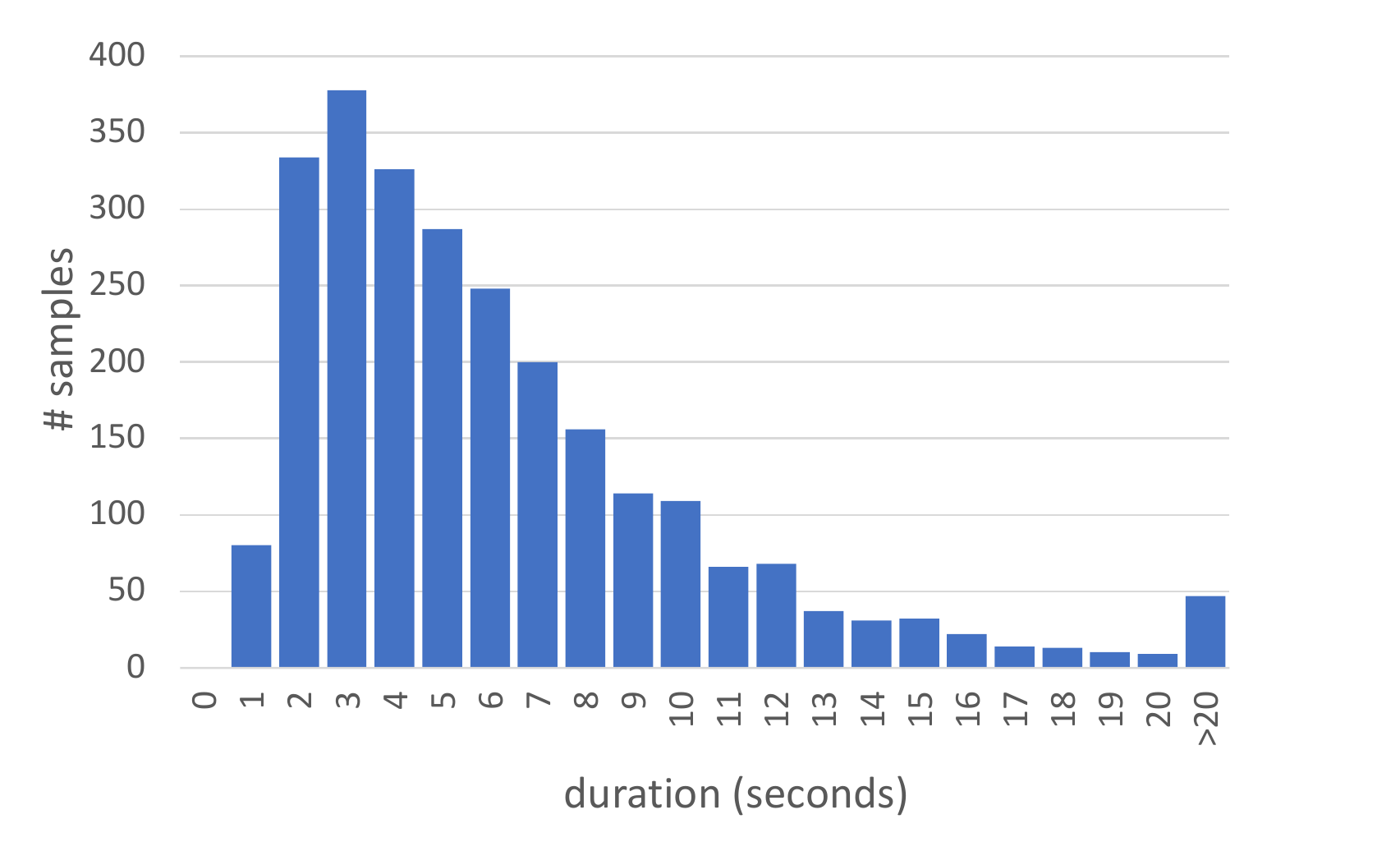}}
\label{subfig:histo_shas_gold}
\\
\subfloat[pDAC segmentation]{\includegraphics[width=9cm]{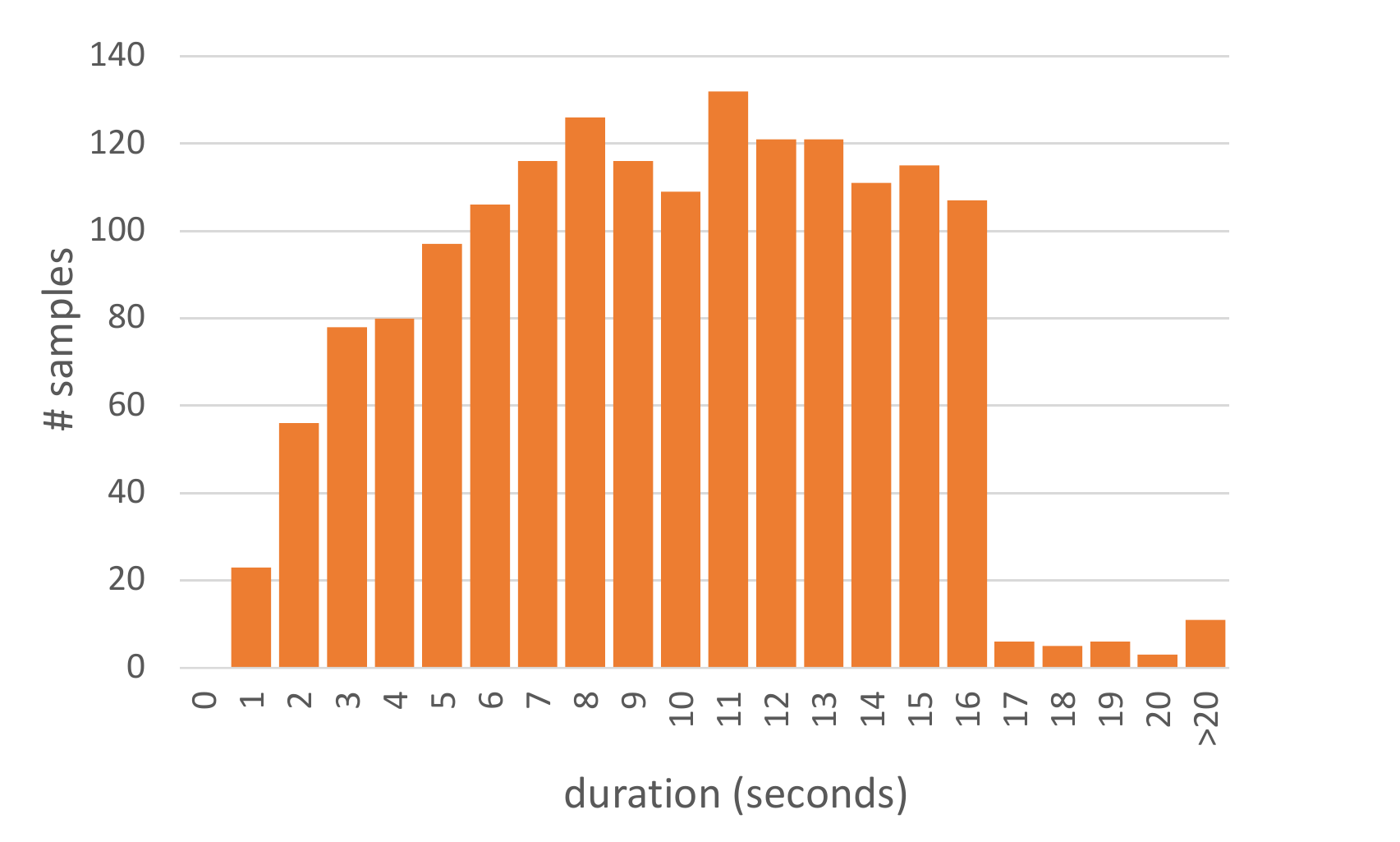}}
\label{subfig:histo_shas_shas}

\caption{Histograms of segment length in each segmentation: Horizontal axis indicates length (seconds) of audio segment, and vertical axis indicates number of samples.}
\label{fig:histo_shas}
\end{figure}

SHAS outperformed the existing pause-based and length-based segmentation methods and consistently achieved better translation quality across multiple language pairs.
However, the segments generated by SHAS tended to be significantly longer than the MuST-C segments.
For example, the average length of the speech segments in MuST-C English-to-German was 5.79 seconds; the average length of segments generated by pDAC was 9.17 seconds.
Fig. \ref{fig:histo_shas} shows the segment length distribution for each segmentation.
The mode value for the sentence-aligned speech segmentation is about three seconds, whereas the mode for the pDAC segmentation is noticeably longer, about 11 seconds.
\par
% Such long segments are caused by pDAC because it stops segmentation once the segment length falls below a predefined value $max$.
The cause of such long segments is that pDAC stops the segmentation once the segment length falls below a predefined value $max$, as mentioned.
% Therefore, while SHAS is a corpus-based method based on SFC predictions, it also has the aspect of a length-based segmentation.
% Therefore, SHAS also has the aspect of a length-based segmentation, while it is a corpus-based method based on SFC predictions.
Thus, although SHAS is a corpus-based segmentation method using SFC, it also has aspects of length-based segmentation.
Its advantage is that speech can be split into segments whose lengths are preferred by ST.
% This approach, which heavily relies on length heuristics, has the advantage that it can split speech into segments of easily translatable length.
Tsiamas et al. \cite{tsiamas22_interspeech} found that a $max$ of values between 14-18 seconds works well, which are considerably longer than the average MuST-C segments' length of 5.79 seconds.
\par
On the other hand, longer segments by SHAS can degrade the translation quality and the time efficiency of ST.
% On the other hand, lengthy segments may degrade translation quality.
% It is known that the longer the segment, the more likely it is that translation omissions will occur.
The longer a segment is, the more likely that translation omissions will occur by neural machine translation \cite{cho-etal-2014-properties}.
% In addition, as segment length increases, the number of samples that can be translated in parallel decreases, and the computation time per token increases due to the increased number of autoregressions.
In addition, the longer a segment is, the more computational time that is required for translation due to the increased time complexity, longer decoder outputs, etc.
%In addition, the longer the segment, the longer the computational time required for translation due to the increased time complexity, the increased number of autoregressions of the decoder, etc.
% Therefore, long segments with SHAS reduce the time efficiency of STs.
% In addition, the longer the segment, the greater the number of autoregressions in the ST model and the longer the computation time required for translation.
% In addition, the longer the segment, the longer the computation time required for translation due to the increased time complexity of the encoder and the number of autoregressions of the decoder, etc.
% In addition, the longer the segment, the longer the computation time required for translation.
% Therefore, long segments with SHAS reduce the time efficiency of STs.
\par
% In order to close the gap between SHAS and gold segmentation, we need a segmentation algorithm that does not heavily rely on length heuristics and improved accuracy of SFC.
To bridge the gap between SHAS and ST corpus segmentation, we need a segmentation algorithm that does not heavily rely on length heuristics.
In this case, SFC predictions become more critical to translation accuracy.
% The reliance on length heuristics raises another problem: it blinds the room for further improvement in translation quality by increasing the accuracy of SFC.
% In other words, pDAC segmentation hides the possibility that length heuristics may not be necessary if SFC accuracy is high enough.

\section{Proposed Method}

Next we propose an online decoding algorithm that focuses on SFC prediction rather than segment length to produce shorter segments (\ref{subsec:pthr}).
We also introduce efficient fine-tuning to update the parameters of the upper layers of wav2vec 2.0 to improve SFC accuracy (\ref{subsec:ft}) \footnote{The source code is available at \url{https://github.com/ahclab/Wav2VecSegmenter}}.

\subsection{Probability-first decoding algorithm with moving average}
\label{subsec:pthr}

\label{algo:thr}
\begin{algorithm}[t]
\caption{pTHR+MA}
\label{algo:pthr}
\begin{algorithmic}[1]
\Inputs{$probs,~max,~min,~thr,~n\_ma,~lerp_{min},~lerp_{max}$}
\Initialize\\
\State{\ind$segments \gets \text{empty List}$}
\State{\ind$start \gets 0$}
\State{\ind$thrs \gets \text{List with size}~max$}
\State{\ind$thrs[:min] \gets 0,~thrs[min:] \gets thr$}\\
\Comment{Set threshold filter $thrs$}
\State{\ind$thrs \gets Lerp(thrs,~min,~lerp_{min},~0,~thr)$}
\State{\ind$thrs \gets Lerp(thrs,~lerp_{max},~max,~thr,~1)$}\\
    \Comment{Apply Linear Interpolation}
\State{\ind$probs \gets MovingAverage(probs,~n\_ma) $}\\
    \Comment{Apply Moving Average of $n\_ma$ frames}
\While{$start < probs.\mathrm{length}$}
    \If{$probs[start] \leq thr$}
        \State{$start \gets start+1$}
    \Else
        \State{$end \gets \mathrm{min}(start+max,~probs.\mathrm{length})$}
        % \State{$ends \gets \text{List of index of } probs[start:start+max][i] \leq thrs[i]$}
        \For{$i = start \, \ldots \, end$}
            \If{$probs[i] \leq thrs[i]$}
                \State{$end \gets i$}
                \State break
            \EndIf
        \EndFor
        % \If{$ends.\mathrm{length} > 0$}
        %     \State{$end \gets start+ends[0]$}
        % \Else
        %     \State{$end \gets \mathrm{min}(start+max,~probs.\mathrm{length})$}
        % \EndIf
        \State{\textbf{append} Tuple($start$,~$end$) to $segments$}
    \EndIf
\EndWhile
\State{return $segments$}
\end{algorithmic}
\end{algorithm}

\begin{figure}[t]
\centering
\includegraphics[width=9cm]{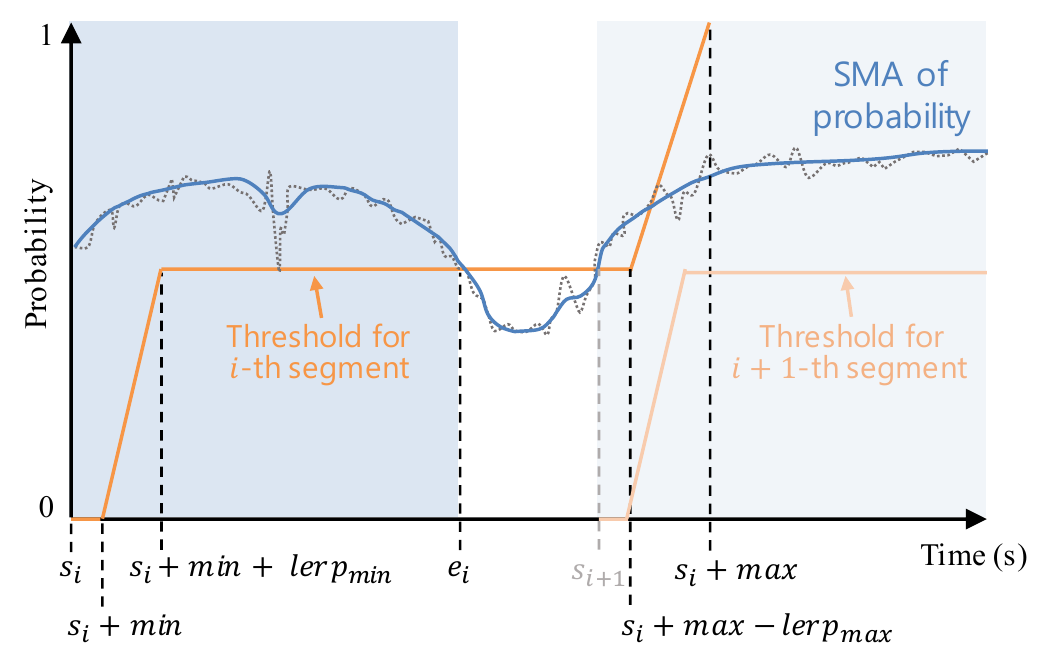}
\caption{Schematic diagram of proposed decoding algorithm.}
\label{fig:pthr}
\end{figure}

% We introduce a segmentation algorithm based on probability thresholds (pTHR).
% pTHR progressively determines where segments start and end using probabilities of each speech frame being included in a segment ($probs$), predicted by SFC.
We introduce a segmentation algorithm based on probability thresholds (pTHR) that uses SFC predictions to find sentence boundaries.
pTHR progressively determines where segments start and end using the probabilities of each speech frame that is included in a segment corresponding to a sentence ($probs$), predicted by SFC.

% The algorithm utilizes three hyperparameters: a threshold to determine if each frame is included in a segment, a maximum segment length ($max$), and a minimum segment length ($min$).
The algorithm simply takes the point at which the probability of being included in a segment exceeds the threshold as starting point $s_i$ of segment $i$ and the point at which it again falls below it as end point $e_i$.
% The algorithm utilizes the following hyperparameters: a threshold filter ($thrs$) to determine if each frame is included in a segment, a maximum segment length ($max$), and a minimum segment length ($min$).
$thrs$ is a sequence of probability thresholds that determine the end of a segment, and its length is the number of speech frames corresponding to $max$ seconds.
The values contained in $thrs$ are almost $thr$ (e.g., 0.5), although they are set to 0 at positions below $min$ to ensure that the segment length is greater than or equal to $min$.
We also applied linear interpolation between $thrs[min:lerp_{min}]$ and $thrs[max-lerp_{min}:]$ to bias the segment lengths to fall within the normal range.
In Algorithm \ref{algo:pthr}, $Lerp(list, start, end, a, b)$ linearly interpolates the values in $list$ between $start$ and $end$, transitioning smoothly from value $a$ at $start$ to value $b$ at $end$.
The segmentation procedure is as follows:
\begin{enumerate}
  \item The algorithm sequentially looks at the $probs$ values, starting with 0. A point at which the value first exceeds the threshold $thr$ is taken as the starting position of the first segment, $s_1$.
  \item $probs[s_1:s_1+max]$ and $thrs$ are compared, and first point $j$, where $probs[j] < thrs[j]$, is taken as the end position of the first segment $e_1$. If no position $j$ is found, $s_1+max$ is set to $e_1$.
  \item A point where the probability exceeds $thr$ again is taken as the starting position of the second segment $s_2$. The positions of the second, third, … segments are identically determined as the first segment.
\end{enumerate}

% The proposed algorithm ensures that the segment length is longer than $min$ and less than or equal to $max$.
% Specifically, $thr$ is set to 0 in the range of $s_i$ to $s_i+min$ and to 1 at $s_i+max$, as shown in Figure \ref{fig:pthr}.
% Linear interpolation (lerp) is also performed in the range of $s_i+min$ to $s_i+min+lerp_{min}$ and $s_i+max-lerp_{max}$ to $s_i+max$.
Many existing VAD methods detect speech segments by such thresholding values such as acoustic power or CTC probabilities, etc.
Our algorithm, pTHR, differs from them by taking SFC predictions as input.
We automate the sentence-level segmentation as given in the ST corpus segmentation instead of performing VAD.

Since pTHR performs thresholding without past information, it can be computed in parallel and at high speed.
On the other hand, pTHR is a less stable method than pDAC because its results are highly dependent on SFC accuracy.
To stabilize the SFC’s prediction, we first tried to use an autoregressive model as SFC, but the model could not be trained due to data imbalance.
Then, inspired by classical time-series analysis methods, we incorporated a simple moving average (SMA) \cite{yule1909applications} into pTHR to smooth the SFC predictions.
We applied an SMA with a window size of $n_{ma}$ frames to $probs$, which are the pTHR inputs.
Specifically, in line 11 of Algorithm \ref{algo:pthr}, $probs[i]$ is updated as follows:
\begin{eqnarray*}
    probs'[i] &=& \frac{1}{n_{ma}} \sum_{k=\max(i-n_{ma},0)}^{i} probs[k] \\
    probs[i] &=& probs'[i]
\end{eqnarray*}
This allows pTHR to stably perform segmentation even when the SFC prediction accuracy is low while maintaining high speed.
We discuss the relationship between SFC prediction accuracy and the stability of each algorithm in Section \ref{subsec:res_algo}.
Hereafter, we refer to our proposed algorithm with $n_{ma}=0$ as pTHR and with $n_{ma}>0$ as pTHR+MA.
Pseudocode and a schematic diagram of the pTHR+MA algorithm are shown in Algorithm \ref{algo:pthr} and Fig. \ref{fig:pthr}.
% In addition, $probs$ is stabilized by applying a simple moving average (SMA) over $n_{ma}$ frames.
% Hereafter, we refer to the proposed algorithm with $n_{ma}=0$ as pTHR and with $n_{ma}>0$ as pTHR+MA.
% Pseudocode and a schematic diagram of pTHR+MA algorithm are shown in Algorithm \ref{algo:pthr} and Figure \ref{fig:pthr}.

\subsection{SFC with memory efficient fine-tuning}
\label{subsec:ft}

\begin{figure}[t]
\centering
\includegraphics[keepaspectratio, width=8.9cm]{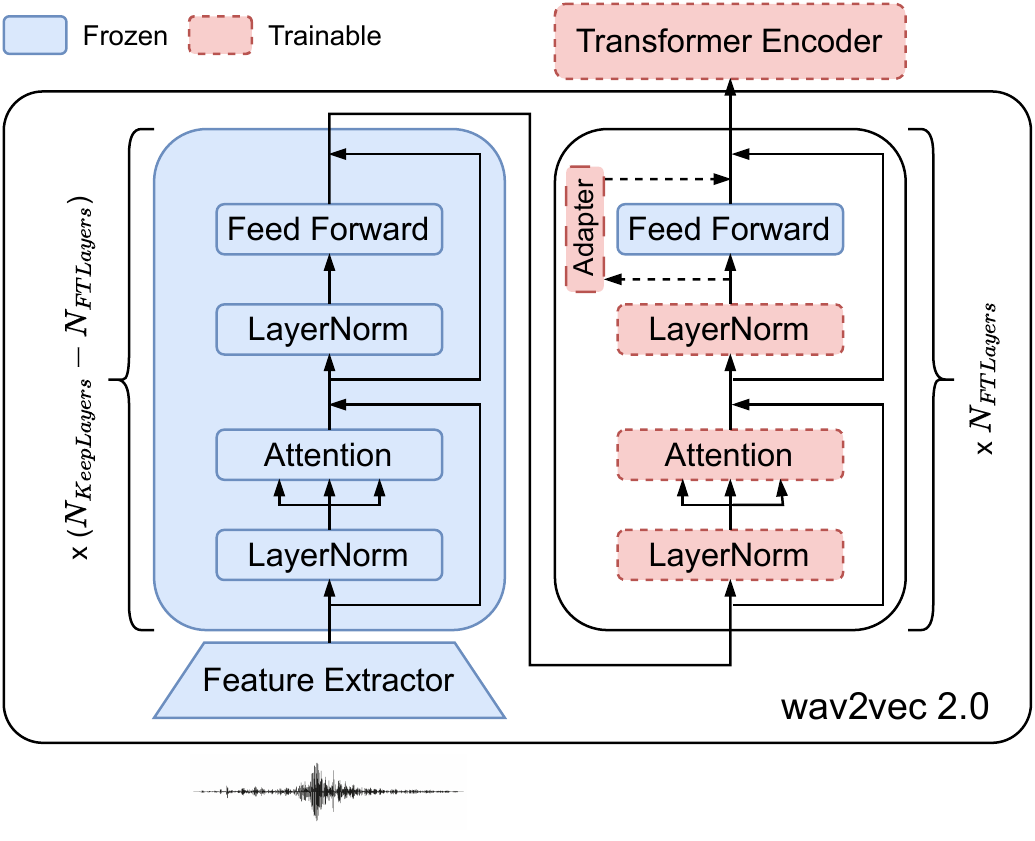}
\caption{Segmentation frame classifier with efficient fine-tuning}
\label{fig:model}
\end{figure}

In our proposed algorithm, the segmentation heavily relies on the SFC prediction, and the translation quality is expected to be affected by SFC accuracy.
In SHAS, the wav2vec 2.0 parameters were frozen during the SFC training.
In contrast, we introduce $SHAS+FTPT$, which updates the wav2vec 2.0 parameters of SFC (Fig. \ref{fig:model}).
The parameters of the upper $N_{FTLayers}$ encoder layers are updated out of $N_{AllLayers}$ layers inherited from the wav2vec 2.0.
% For memory efficiency, the parameters of the feed-forward layers are frozen, and a parallel adapter~\cite{he2021towards} is added ~\cite{tsiamas-etal-2022-pretrained} to substitute for the feed-forward layers.
% \red{
% For memory efficiency, the parameters of the feed-forward layers are frozen, and parallel adapters ~\cite{he2021towards} are added to substitute for the feed-forward layers. Parallel adapters, which are shown by He et al. \cite{he2021towards} to achieve higher performance than adapters inserted sequentially, have also been verified for their effectiveness in ST \cite{tsiamas-etal-2022-pretrained}.
% }
While the self-attention mechanism has quadratic complexity with respect to input length, our model operates with fixed-length inputs during both training and inference, ensuring consistent memory allocation for self-attention. To further optimize memory usage, we froze the parameters of the feed-forward layers, which can be substantial in terms of parameter count. In their place, we introduced parallel adapters ~\cite{he2021towards}. These parallel adapters, as demonstrated by He et al. \cite{he2021towards}, outperform sequentially inserted adapters and have been validated for their efficacy in ST \cite{tsiamas-etal-2022-pretrained}.

\section{Experimental Settings}
\label{sec:exp}
% In order to investigate the effectiveness of the proposed method, we conducted speech translation experiments and compared several speech segmentation methods.
We investigated the effectiveness of our proposed method by conducting speech translation experiments and compared several speech segmentation methods.

\subsection{Data}
\label{subsec:exp_data}

% \begin{table}[!t]
% \caption{Number of segments of MuST-C v2 used in the experiments.}
% \label{tab:statistics}
% \centering
% \begin{tabular}{|l||c|c|c|}
% \hline
% Language pair & train & dev & tst-COMMON \\ \hline \hline
% English-German & 250,942 & 1,415 & 2,580 \\
% English-Japanese & 328,639 & 1,369 & 2,841 \\ \hline
% \end{tabular}
% \end{table}
\begin{table}[!t]
\caption{Number of segments of MuST-C v2 used in experiments}
\label{tab:statistics}
\centering
\begin{tabular}{|l||c|c|c|}
\hline
Language pair & train & dev & tst-COMMON \\ \hline \hline
English-to-German & 250,942 & 1,415 & 2,580 \\ \hline
\end{tabular}
\end{table}
% \begin{table}[!t]
% \caption{Dataset statistics of MuST-C v2 used in the experiments.}
% \label{tab:statistics}
% \centering
% \begin{tabular}{|l||cc|c|}
% \hline
% \multirow{2}{*}{Dataset} & \multicolumn{2}{c|}{Training} & Test \\ \cline{2-4} 
%  & \multicolumn{1}{c|}{\# segments} & \# hours & \# segments \\ \hline \hline
% MuST-C v2 En-De & \multicolumn{1}{c|}{229,703} & 436 & 2,641 \\ \hline
% mTEDx & \multicolumn{1}{c|}{} &  &  \\
% ~Es-En & \multicolumn{1}{c|}{36,263} & 69 & 996 \\
% ~Fr-En & \multicolumn{1}{c|}{30,171} & 50 & 1,041 \\
% ~It-En & \multicolumn{1}{c|}{24,576} & 53 & 979 \\
% ~Pt-En & \multicolumn{1}{c|}{30,855} & 59 & 1,022 \\ \hline
% Europarl ST En-De & \multicolumn{1}{c|}{} & 83 &  \\ \hlinerr
% CoVoST 2 En-De & \multicolumn{1}{c|}{} & 413 &  \\ \hline
% \end{tabular}
% \end{table}

% We conducted experiments with English to German ST.
We conducted experiments with the English-to-German (En-De) ST as our primary forcus.
We used MuST-C v2 for the experiments, which consisted of triplets of segmented English speech, transcripts, and target language translations.
Table \ref{tab:statistics} shows the statistics of the datasets used in the experiments.
MuST-C train and dev split were used to build the SFC models.
tst-COMMON was used as a test set for evaluation.
The average SNRs of the tst-COMMOM are 37dB for the period including only ambient noise only and 1.52dB for the period including applause and laughter.
\par
To further validate the effectiveness of the SFC models across different languages, we performed supplementary tests using the 8 language pairs available in MuST-C v1. Specifically, these tests were conducted from English to German, Spanish (En-Es), French (En-Fr), Italian (En-It), Dutch (En-Nl), Portuguese (En-Pt), Romanian (En-Ro), and Russian (En-Ru).
Additionally, to assess the applicability of the method in different domains, we performed tests using the Europarl-ST En-De dataset \cite{iranzo2020europarl}.
%The audio files of the tst-COMMON contain a lot of applause and laughter, resulting in a low signal-to-noise ratio (SNR) of 1.52.
%\footnote{SNR was calculated as $10 \log ((\sqrt(p_x)-\sqrt(p_n)) / \sqrt(p_n))$, where $p_x$ is the average power of all segments and $p_n$ is the average power outside the segments.}

\subsection{Evaluation}
\label{subsec:exp_eval}

The evaluation process followed \cite{matusov2005evaluating}.
First, the test set audio files were split using one of the segmentation methods (described in \ref{subsec:exp_seg}).
Then the newly created segments were translated using an ST model (\ref{subsec:exp_st}), and the translations were aligned to the references in the test set using mwerSegmenter \cite{matusov2005evaluating}.
Finally, the BLEU scores \cite{papineni-etal-2002-bleu} were calculated with SacreBLEU \cite{post-2018-call}\footnote{\url{https://github.com/mjpost/sacrebleu}}\footnote{signature: BLEU+case.mixed+numrefs.1+smooth.exp+tok.13a+version.1.5.0}.
We also measured BERTScore \cite{zhang2019bertscore}\footnote{bert-base-multilingual-cased} and BLEURT \cite{sellam-etal-2020-bleurt}\footnote{https://storage.googleapis.com/bleurt-oss-21/BLEURT-20.zip}.

\subsection{Segmentation method}
\label{subsec:exp_seg}
% wav2vec 2.0: the 300m parameter model with 24 layers and 1024 dims Transformer encoder: one encoder layers
\subsubsection{SFC}
\label{subsubsec:exp_sfc}
We trained the SFC models with random segments of 20 seconds of audio samples extracted from the training data, following Tsiamas et al. \cite{tsiamas22_interspeech}.
% As SFC input, we used acoustic features of 80-dimensional log mel-filterbank computed every 10ms with a 25ms window shift.
As a pre-trained speech encoder of wav2vec 2.0, we used an XLS-R model \cite{babu22_interspeech} of 300 million parameters\footnote{\url{https://huggingface.co/facebook/wav2vec2-xls-r-300m}}, with 24 layers and a dimensionality of 1024.
The Transformer encoder has a single layer, 1024 model dimensions, 2048 feed-forward dimensions, eight heads, pre-layer normalization, GELU activation, and 0.1 dropout. Prior to being mapped to probabilities through a linear sigmoid layer, an additional layer of normalization and 0.1 dropout were applied. Models were trained for 16 epochs using Adam with an initial learning rate of $2.5 \cdot 10^{-4}$ (decayed with cosine annealing). After training, the best checkpoint was selected based on the prediction performance of the dev set.
\par
As SFCs of the baseline method SHAS (these models are also called $SHAS$), we built a $middle$ model that inherited the lower 16 layers of the XLS-R encoder and a $large$ model that inherited 24 layers.
In their preliminary experiments, Tsiamas et al. \cite{tsiamas22_interspeech} found that it is beneficial to inherit the lower 14 layers from XLS-R.
We also quoted the scores they reported for comparison.
\par
We created the following six variations of $SHAS+FTPT$ shown in Section \ref{subsec:ft}.
The model settings are shown in brackets in the format $N_{FTLayers}$/$N_{AllLayers}$.
\begin{itemize}
    \item \textit{middle+quarter} (4/16)
    \item \textit{middle+half} (8/16)
    \item \textit{middle+all} (16/16)
    \item \textit{large+quarter} (6/24)
    \item \textit{large+half} (12/24)
    \item \textit{large+all} (24/24)
\end{itemize}

\subsubsection{Segmentation algorithm}
\label{subsubsec:exp_algo}

We used pDAC (Section \ref{subsec:pdac}) and pSTRM~\cite{gaido-etal-2021-beyond} as baseline segmentation algorithms.
pSTRM splits on the longest pause in the interval ($min$ and $max$), if any, and otherwise it splits at $max$.
pSTRM also emphasizes the target segments’ length like pDAC, although it is an online decoding algorithm like our proposed algorithms.
The proposed algorithms are pTHR+MA with moving average and pTHR without it.
We tuned hyperparameters of each segmentation algorithm using dev set.
For pDAC and pSTRM, both of which prioritize segment length, we fixed the threshold at 0.5, $min=0.2$, and tried $max=\{28,26,24,22,20,18,16,14,12,10\}$.
For pTHR and pTHR+MA, we fixed $max=28$, $min=0.2$, and tried $thr=\{0.9,0.8,0.7,0.6,0.5,0.4,0.3,0.2,0.1\}$ and a moving average of $\{0,0.1,0.2,0.4,0.8,1\}$ seconds.
As per the above settings, our proposed algorithm also imposes length constraints with $max$ and $min$. However, these are merely safeguards to avoid extreme lengths, and usually, the segmentation positions are determined by probability and threshold $thr$.

% For pDAC, we tuned the max_segment_length in the range of (10,28) seconds
% For pSTREAM, .. For pTHR, ..
% ^* For pSTRM and pDAC, which prioritize length, we fixed the threshold of the model and searched for max_segment_length. For pTHR, which prioritizes model probability, max_segment_length was fixed and threshold was searched.

\subsection{Speech translation model}
\label{subsec:exp_st}

Following Tsiamas et al. \cite{tsiamas22_interspeech}, we used the joint speech-to-text model \cite{tang-etal-2021-improving} from fairseq \cite{wang-etal-2020-fairseq} for MuST-C v2 En-De and Europarl-ST En-De.
This joint model is a Transformer encoder-decoder that can take both speech and text as input and share the top layer of the encoder between the two modalities. It performs knowledge distillation from the text-to-text translation task as a guide for the ST task \cite{liu2019end,gaido2020knowledge} and applies cross-attention regularization to the encoder representations to bridge the gap between the two modalities. We used a model trained on MuST-C En-De\footnote{\url{https://github.com/facebookresearch/fairseq/blob/main/examples/speech_text_joint_to_text/docs/ende-mustc.md}}. This model has 12 encoder and 6 decoder layers, with a dimensionality of 512 and 2048 feedforward dimensions.
For the tests on 8 language pairs using MuST-C v1, we employed a multilingual ST model trained on the MuST-C v1\footnote{\url{https://github.com/facebookresearch/fairseq/blob/main/examples/speech_to_text/docs/mustc_example.md}}.
This model also has 12 encoder and 6 decoder layers, with a dimensionality of 512 and 2048 feedforward dimensions.
During inference, decoding was performed with a beam search of beam size 5.

\section{Experimental Results}
\label{sec:res}

\subsection{Translation quality}
\label{subsec:res_overall}

\begin{table*}[!t]
\caption{Results by baseline model $SHAS$ and proposed model $SHAS+FTPT$, and four decoding algorithms on MuST-C v2 En-De. Numbers in brackets are the number of encoder layers (fine-tuned/inherited from wav2vec 2.0). For BLEU, \dag and \ddag indicate statistical significance ($p<0.05$ and $p<0.001$, respectively) in comparison with the top row.}
\label{tab:unified_results}
\centering
\begin{tabular}{|l|c|c|cc||c|c|cc||c|c|cc|}
\hline
Model~\textbackslash~Decoding & \multicolumn{4}{c||}{BLEU} & \multicolumn{4}{c||}{BERTScore F1} & \multicolumn{4}{c|}{BLEURT} \\
\cline{2-13}
 & pDAC & pSTRM & pTHR & +MA & pDAC & pSTRM & pTHR & +MA & pDAC & pSTRM & pTHR & +MA \\ 
\hline \hline
$SHAS$ &&&&&&&&&&&& \\
~\textit{middle} (0/16) & 25.42 & 25.11 & 23.54 & 25.73 & 0.5201 & 0.5233 & 0.5324 & 0.5418 & 0.4958 & 0.4941 & 0.4992 & 0.5050 \\
~\textit{large} (0/24)  & 24.41 & 25.18 & 21.15 & 24.78 & 0.5072 & 0.5378 & 0.5087 & 0.5381 & 0.4850 & 0.4975 & 0.4730 & 0.4988 \\ 
\hline
$SHAS+FTPT$ &&&&&&&&&&&& \\
~\textit{middle+quarter} (4/16) & 25.84 & 25.57\dag & 25.67\ddag & 25.96 & 0.5381 & 0.5342 & 0.5641 & 0.5592 & 0.5082 & 0.5035 & 0.5232 & 0.5213 \\
~\textit{middle+half} (8/16) & 25.75 & 25.52 & 25.92\ddag & 26.17 & 0.5344 & 0.5343 & 0.5724 & \underline{0.5703} & 0.5046 & 0.5046 & 0.5287 & \underline{0.5267} \\
~\textit{middle+all} (16/16) & 25.73 & 25.71\dag & 26.13\ddag & 26.27\dag & \underline{0.5394} & 0.5401 & 0.5697 & 0.5634 & 0.5054 & 0.5029 & 0.5264 & 0.5211 \\
~\textit{large+quarter} (6/24) & 25.73 & \underline{25.74}\dag & 25.73\ddag & 26.18 & 0.5369 & \underline{0.5420} & 0.5651 & 0.5560 & 0.5058 & \underline{0.5054} & 0.5238 & 0.5183 \\
~\textit{large+half} (12/24) & 25.89\dag & 25.58 & 26.26\ddag & 26.15 & 0.5363 & 0.5358 & \underline{\textbf{0.5751}} & 0.5623 & \underline{0.5077} & 0.5049 & \underline{\textbf{0.5317}} & 0.5205 \\
~\textit{large+all} (24/24) & \underline{25.95}\dag & 25.70\dag & \underline{26.28}\ddag & \underline{\textbf{26.30}\dag} & 0.5518 & 0.5345 & 0.5657 & 0.5698 & 0.5161 & 0.5007 & 0.5239 & 0.5257 \\
\hline
\end{tabular}
\end{table*}

\begin{table}[!t]
\caption{BLEU scores of sentence-aligned segmentation, SHAS, and our method in English-to-German Translation. Numbers in parentheses are the percentages of retained BLEU scores of sentence-aligned speech segmentation}.
\label{tab:gold_shas_proposal}
\centering
\begin{tabular}{|l||c|}
\hline
      & MuST-C v2 En-De         \\ \hline \hline
Sentence-aligned       & 26.99 (100\%) \\ \hline
SHAS (Tsiamas+22) & 25.67 (95.1\%)  \\ \hline
Proposed method        & \textbf{26.30 (97.4\%)}  \\ \hline
\end{tabular}
\end{table}

% Table \ref{tab:main_results} shows the overall results of MuST-C v2 En-De.
% From the perspective of the SFC model, $SHAS+FTPT$ generally achieved higher translation accuracy than $SHAS$. In terms of algorithms, when using $SHAS+FTPT$ for SFC, pTHR and pTHR+MA tended to be either comparable to or even better than the baseline.
% The best BLEU score (26.30) was obtained when using the $large+all$ of the $SHAS+FTPT$ for SFC and pTHR+MA for the segmentation algorithm.
% We discuss the effects of the proposed segmentation algorithm in Section \ref{subsec:res_algo} and fine-tuning the wav2vec 2.0 in Section \ref{subsec:res_ft}.
% Table \ref{tab:bertscore_results} shows the results for BERTScore F1. While the trend of the results was similar to the previously discussed BLEU scores, differences between our proposed algorithms and the baselines were more pronounced. This could potentially be attributed to the longer segments produced by the baseline algorithms. Specifically, as the translation segments become longer, there is an increased risk of misalignment during BERTScore computation, which can lead to a decrease in the score.
% Table \ref{tab:bleurt_results} presents the BLEURT results, which showed a similar trend to BERTScore, suggesting that shorter segments might receive higher evaluations in embedding-based automatic evaluations.

Table \ref{tab:unified_results} shows the overall results of MuST-C v2 En-De across BLEU, BERTScore F1, and BLEURT metrics.
The leftmost columns in the table present the BLEU results.
From the perspective of the SFC model, $SHAS+FTPT$ generally achieved higher translation accuracy than $SHAS$. In terms of algorithms, when using $SHAS+FTPT$ for SFC, pTHR and pTHR+MA tended to be either comparable to or even better than the baseline.
The best BLEU score (26.30) was obtained when using the $large+all$ of the $SHAS+FTPT$ for SFC and pTHR+MA for the segmentation algorithm.
We discuss the effects of the proposed segmentation algorithm in Section \ref{subsec:res_algo} and fine-tuning the wav2vec 2.0 in Section \ref{subsec:res_ft}.
The middle columns in the table display the results for BERTScore F1. While the trend of the results was similar to the previously discussed BLEU scores, differences between our proposed algorithms and the baselines were more pronounced. This could potentially be attributed to the longer segments produced by the baseline algorithms. Specifically, as the translation segments become longer, there is an increased risk of misalignment during BERTScore computation, which can lead to a decrease in the score.
The rightmost columns of Table \ref{tab:unified_results} present the BLEURT results.
These showed a similar trend to BERTScore, suggesting that shorter segments might receive higher evaluations in embedding-based automatic evaluations.

Table \ref{tab:gold_shas_proposal} shows the translation qualities of the segmentations of MuST-C, SHAS, and the proposed method.
The SHAS's score is that reported by Tsiamas et al., and the proposed method's score is the best result from Table \ref{tab:unified_results}.
The proposed method retained 97.4\% of the BLEU score for sentence-aligned speech segmentation, surpassing the 95.1\% by SHAS.

\subsection{Effectiveness of segmentation algorithm}
\label{subsec:res_algo}

% \begin{figure}[!t]
% \centering
% \subfloat[Middle-sized models.]{\includegraphics[width=8.9cm]{figures/param_bleu_middle_bar_2.pdf}}
% \\
% \centering
% \subfloat[Large-sized models.]{\includegraphics[width=8.9cm]{figures/param_bleu_large_bar_2.pdf}}
% \caption{BLEU for each segmentation algorithm with different segmentation models}
% \label{fig:parameter_bleu}
% \end{figure}

% \sout{Figure \ref{fig:parameter_bleu} compares of the four segmentation algorithms in (a) four middle-sized models ($middle$, $middle+quarter$, $middle+half$, $middle+all$) and (b) four large-sized models ($large$, $large+quarter$, $large+half$, $large+all$).}
In table \ref{tab:unified_results}, for the baseline SFC models $SHAS$ trained with fixed wav2vec 2.0 parameters ($middle$ and $large$), the pTHR results had significantly lower BLEU than those of the conventional segmentation algorithms, pDAC and pSTRM.
This result implies that the SFC performance that predicted the ST corpus segmentation was insufficient, and in such cases, the conventional segmentation algorithms, which heavily rely on length heuristics, had an advantage.
% On the other hand, as the number of layers to be trained $N_{FTLayers}$ increased, the pTHR results improved significantly over pDAC and pSTRM.
% Specifically, only pTHR showed a statistically significant difference between $SHAS$ and $SHAS+FTPT$ at $p<0.001$, while some other algorithms showed significance at $p<0.05$.
% On the other hand, as the number of layers trained $N_{FTLayers}$ increased, the performance of pTHR markedly improved, achieving BLEU scores that were comparable to those of pDAC and pSTRM.
On the other hand, as the number of layers to be trained $N_{FTLayers}$ increased, the pTHR results improved, achieving BLEU scores that were comparable to those of pDAC and pSTRM.
While other algorithms demonstrated a statistically significant difference between $SHAS$ and $SHAS+FTPT$ at a $p<0.05$ level, only pTHR exhibited significance at $p<0.001$, underscoring its substantial improvement.
It suggests that the need to consider segment length decreases with higher SFC accuracy.
\par
Moreover, pTHR with a moving average (pTHR+MA) obtained the best BLEU score in most models.
In particular, for a \textit{large} model, it outperformed pTHR by more than 3 BLEU points, demonstrating the effect of smoothing the probability using the moving average to compensate for the model's low prediction accuracy.
However, we found no significant difference between pTHR and pTHR+MA for $large+half$, $large+all$, etc., where there are many trainable parameters.
The best parameters for each segmentation algorithm are shown in Appendix \ref{app:best_parameters}.

\subsection{Effectiveness of wav2vec 2.0 fine-tuning}
\label{subsec:res_ft}

\begin{table}[!t]
\caption{Number of trainable parameters and maximum GPU memory usage for each SFC}
\label{tab:num_params}
\centering
\begin{tabular}{|l|c|c|}
\hline
Model & \multicolumn{1}{c|}{\begin{tabular}[c]{@{}c@{}}Trainable~/~Non-trainable\\ parameters\end{tabular}} & \multicolumn{1}{c|}{\begin{tabular}[c]{@{}c@{}}GPU memory\\ (MB)\end{tabular}} \\ \hline \hline
~\textit{middle} (0/16) & 8M~/~215M & 4,469 \\
~\textit{middle+quarter} (4/16) & 38M~/~189M & 15,511 \\
~\textit{middle+half} (8/16) & 59M~/~173M & 15,877 \\
~\textit{middle+all} (16/16) & 101M~/~139M & 17,053 \\ \hline
~\textit{large} (0/24) & 8M~/~315M & 5,716 \\
~\textit{large+quarter} (6/24) & 48M~/~282M & 21,570 \\
~\textit{large+half} (12/24) & 80M~/~257M & 23,058 \\
~\textit{large+all} (24/24) & 143M~/~206M & 25,272 \\ \hline
\end{tabular}
\end{table}

% [TODO] change this to one drawn by Excel
% [TODO] Add a figure for \textit{last}
\begin{figure}[!t]
\centering
\includegraphics[width=8.9cm]{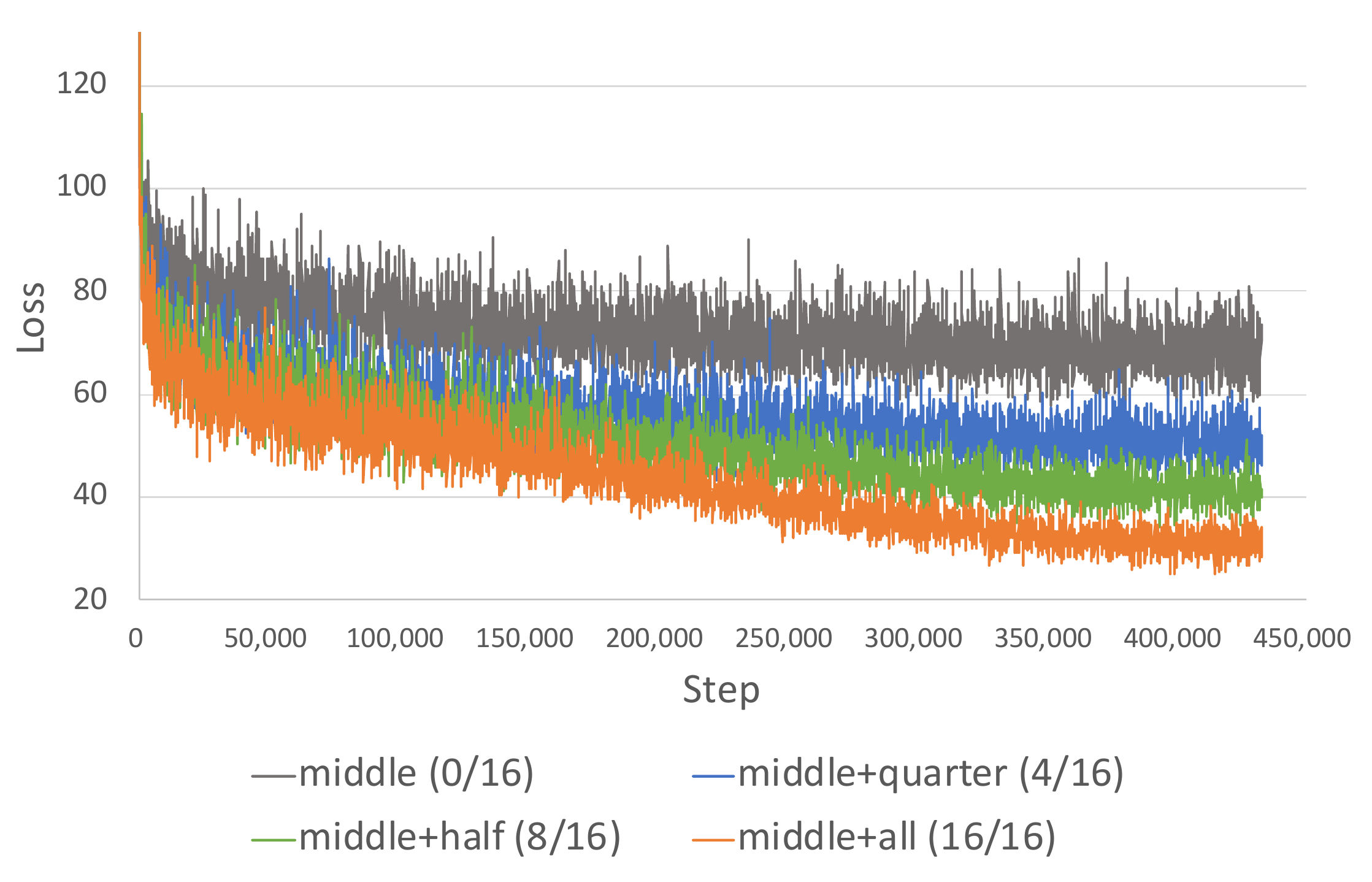}
\caption{Loss curves of middle SFC models}
\label{fig:loss_curve}
\end{figure}

\begin{table}[!t]
\caption{Prediction performance for each SFC model}
\label{tab:sfc_acc}
\centering
\begin{tabular}{|l|c|c|c|}
\hline
Model & Precision & Recall & F1 \\ \hline \hline
~\textit{middle} (0/16) & 0.9894 & 0.9046 & 0.9449 \\
~\textit{middle+quarter} (4/16) & 0.9879 & 0.9194 & 0.9524 \\
~\textit{middle+half} (8/16) & 0.9861 & 0.9282 & 0.9563 \\
~\textit{middle+all} (16/16) & 0.9834 & 0.9344 & 0.9583 \\ \hline
~\textit{large} (0/24) & 0.9802 & 0.8532 & 0.9123 \\
~\textit{large+quarter} (6/24) & 0.9908 & 0.9074 & 0.9472 \\
~\textit{large+half} (12/24) & 0.9896 & 0.9166 & 0.9517 \\
~\textit{large+all} (24/24) & 0.9812 & 0.9381 & 0.9591 \\ \hline
\end{tabular}
\end{table}

Table \ref{tab:num_params} shows the number of trainable and non-trainable parameters and the maximum GPU memory usage for each SFC.
The translation quality by pTHR is somewhat proportional to the number of trainable parameters in SFC.
The explanation is that the higher the percentage of parameters that can be updated, the easier it is to fit a pre-trained speech model to the segmentation task, as shown by the loss curves in Fig. \ref{fig:loss_curve}.
Table \ref{tab:sfc_acc} shows the prediction performance for the dev set by each SFC model.
After determining the output to be 0 or 1 with a threshold value of 0.5 from the probability for each frame, we calculated the Precision, Recall, and F1 for the correct label.
$SHAS$ models ($middle$ and $large$) have a high Precision of about 98\%, but a low Recall of about 85\% to 90\%.
Label $y = 1$ indicates that the corresponding frame is inside the segment, while $y = 0$ indicates that it is outside of it.
Therefore, low Recall shows that in-segment frames are often incorrectly judged as out-of-segment.
In SHAS, the length heuristics with pDAC mitigated the over-segmentation due to this low Recall.
On the other hand, $SHAS+FTPT$ models ($middle+all$, $large+all$), with fine-tuning of all the layers, improved the Recall by 3\% to 7\% with almost no drop in Precision.
The reduction in the need for length heuristics proportional to model size, mentioned in Section \ref{subsec:res_algo}, can be explained by this improvement in Recall.
\subsection{Improving time efficiency of ST}

\begin{figure}[!t]
\centering
\includegraphics[width=8.9cm]{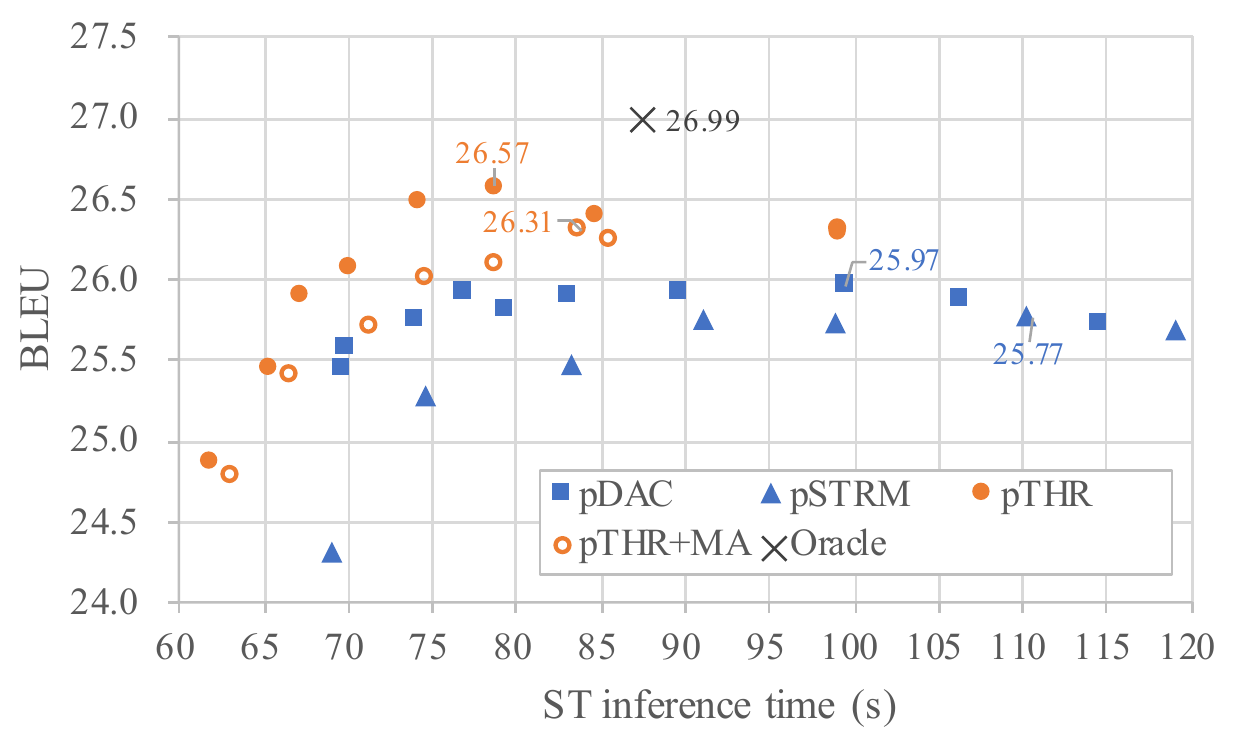}
\caption{ST inference time and BLEU for each segmentation method. (model=large+all)}
\label{fig:time_efficiency}
\end{figure}

Figure \ref{fig:time_efficiency} shows the trade-off between time efficiency and translation quality for each segmentation algorithm with SFC model $large+all$.
The number of tokens per mini-batch was set to 100,000, and ST inference was performed using NVIDIA GeForce RTX 3090 on the same computer.
Segments were sorted by length before batching.
The horizontal axis shows the average ST inference times of five inferences, and the vertical axis shows the BLEU.
For pDAC and pSTRM, the conditions were set in the range of $max=[2,28]$, and for pTHR and pTHR+MA, they were set in the range of $threshold=[0.1,0.9]$.
The proposed algorithms (pTHR and pTHR+MA) achieved higher translation accuracy with better time efficiency than the baseline algorithms (pDAC and pSTRM).
In particular, the segments generated by pTHR were processed about 25\% faster than the pDAC segments while retaining about 97\% of the translation accuracy of the sentence-aligned speech segmentation.
The shorter average segment length of 5.67 seconds for pTHR compared to 9.17 seconds for pDAC contributed to the higher speed because shorter segment lengths are easier to parallelize and reduce the number of autoregressions.

\subsection{Segment length distribution}

% \begin{figure*}[ht]
%     \begin{tabular}{cc}
%       \begin{minipage}[t]{0.45\hsize}
%         \centering
%         \includegraphics[keepaspectratio, width=8.5cm]{figures/histogram_pstrm_noscore.pdf}
%         \subcaption{pSTRM segmentation}
%       \end{minipage} &
%       \begin{minipage}[t]{0.45\hsize}
%         \centering
%         \includegraphics[keepaspectratio, width=8.5cm]{figures/histogram_pthr_noscore.pdf}
%         \subcaption{pTHR segmentation}
%       \end{minipage} \\
%     \begin{minipage}[t]{0.45\hsize}
%         \centering
%         \includegraphics[keepaspectratio, width=8.5cm]{figures/histogram_pthr_ma_noscore.pdf}
%         \subcaption{pTHR+MA segmentation}
%       \end{minipage} 
%     \end{tabular}
%     \caption{Histograms of segment length in each segmentation}
%     \label{fig:histogram}
% \end{figure*}

\begin{figure*}[ht]
    \centering
    \begin{tabular}{ccc}
      \begin{minipage}[t]{0.31\hsize}
        \centering
        \includegraphics[keepaspectratio, width=5.6cm]{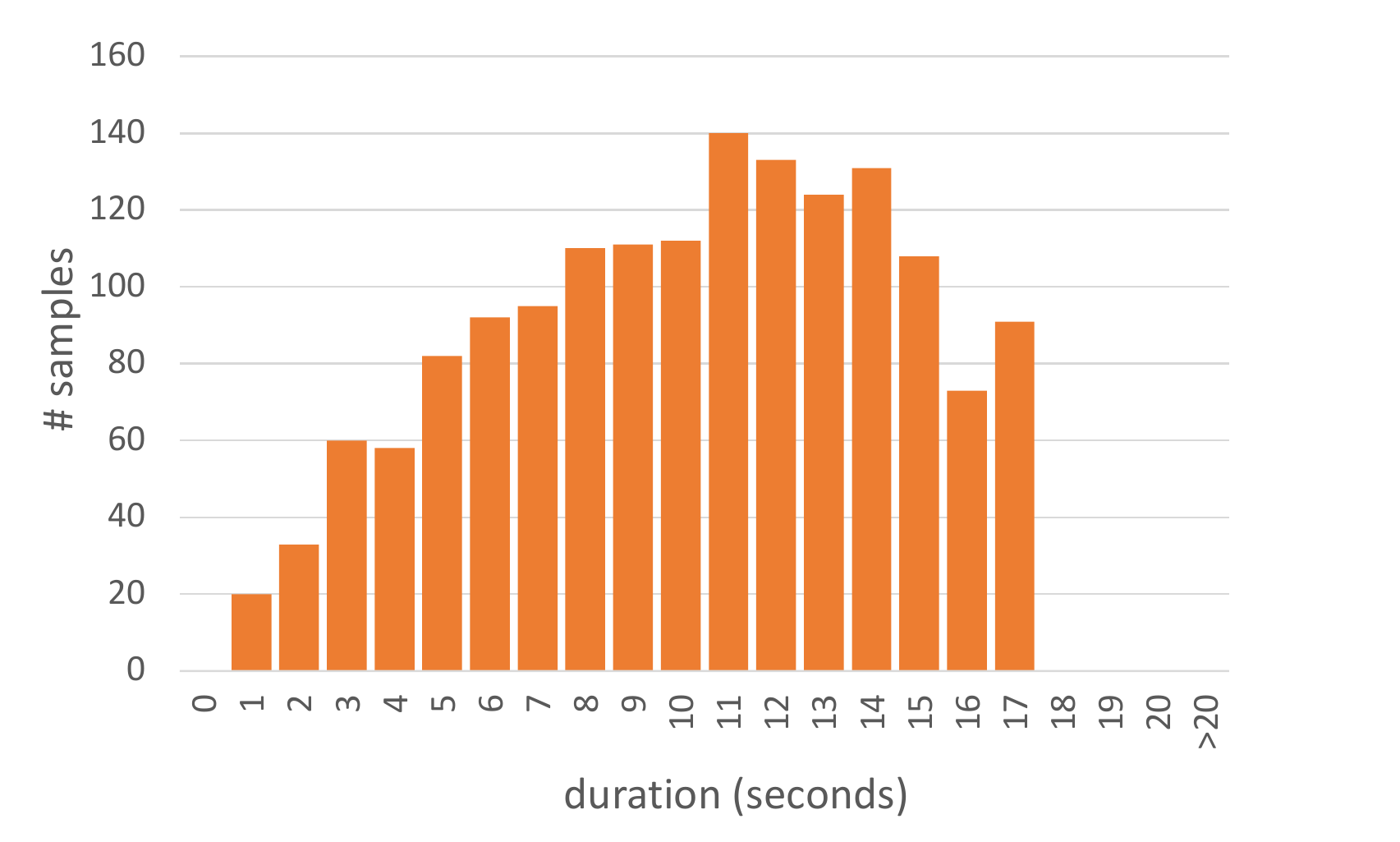}
        \subcaption{pSTRM segmentation}
      \end{minipage} &
      \begin{minipage}[t]{0.31\hsize}
        \centering
        \includegraphics[keepaspectratio, width=5.6cm]{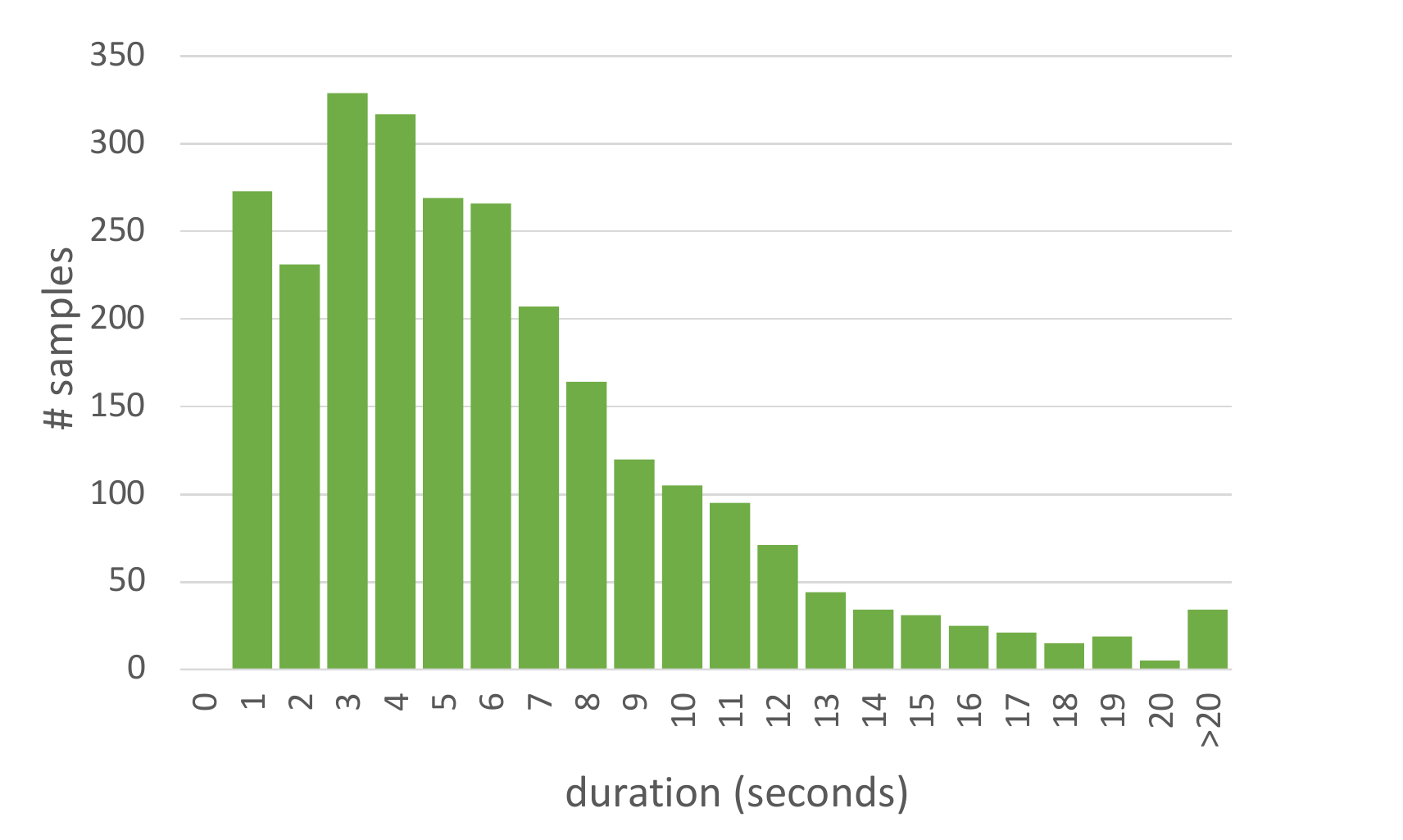}
        \subcaption{pTHR segmentation}
      \end{minipage} &
      \begin{minipage}[t]{0.31\hsize}
        \centering
        \includegraphics[keepaspectratio, width=5.6cm]{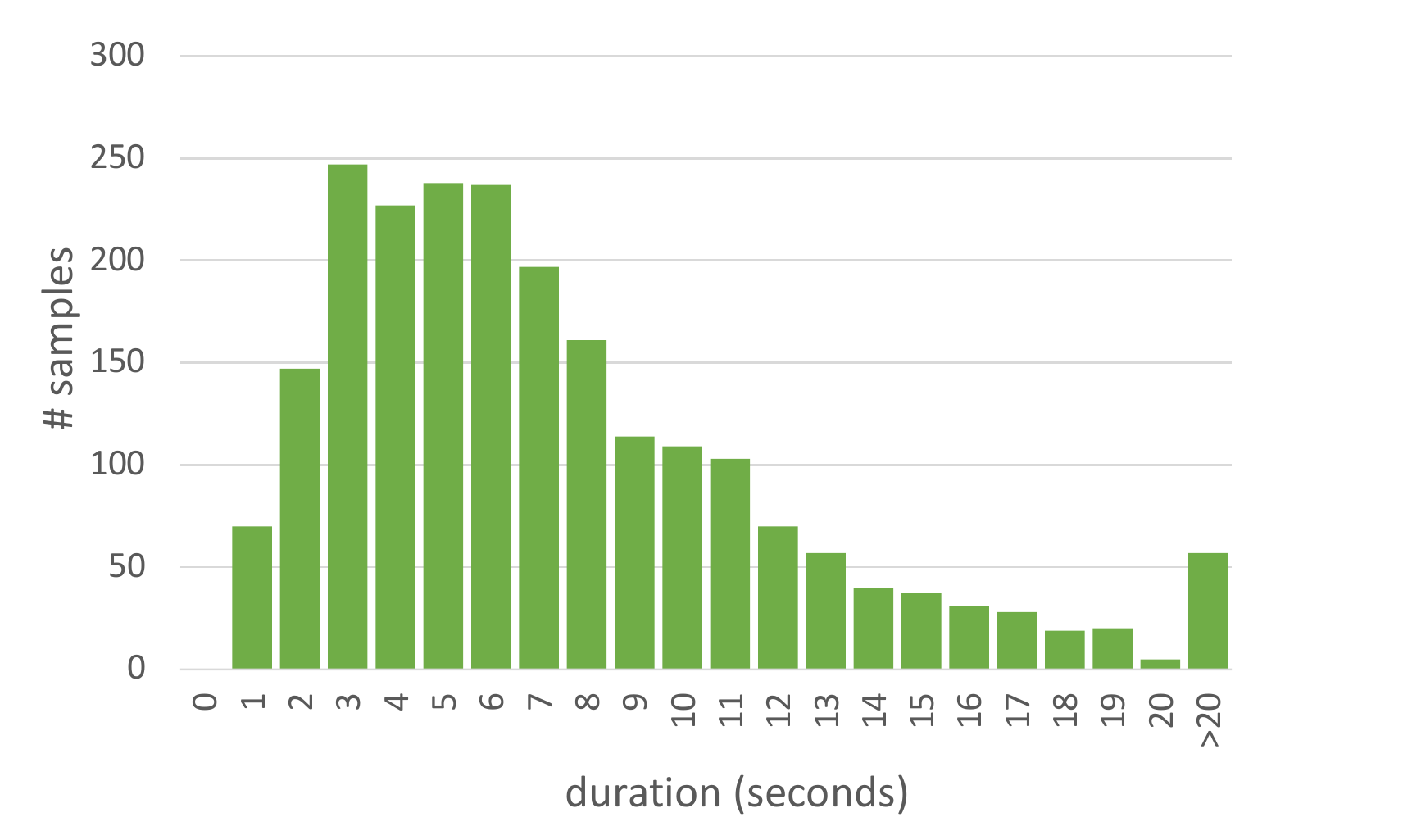}
        \subcaption{pTHR+MA segmentation}
      \end{minipage}
    \end{tabular}
    \caption{Histograms of segment length in each segmentation}
    \label{fig:histogram}
\end{figure*}

Figure \ref{fig:histogram} shows the length distribution of the segments generated by pSTRM, pTHR, and pTHR+MA using an SFC model $large+all$.
% pDAC tends to generate longer segments, as shown in Section \ref{subsec:problem}; the same is true for pSTRM.
pDAC (Fig. \ref{fig:histo_shas}b) tends to produce longer segments compared to the sentence-aligned speech segmentation (Fig. \ref{fig:histo_shas}a), as shown in Section \ref{subsec:problem}; the same is true for pSTRM (Fig. \ref{fig:histogram}a).
On the other hand, the pTHR (Fig. \ref{fig:histogram}b) and pTHR+MA (Fig. \ref{fig:histogram}c) distributions resemble that of the sentence-aligned speech segmentation, producing relatively short segments, suggesting that the segmentation of the proposed method was more faithful to sentence segmentation than SHAS.
\par
In another aspect, longer pDAC and pSTRM segments reduced the contextual dependencies between segments, perhaps improving the translation accuracy.
Therefore, combining pTHR and pTHR+MA with a context-aware ST \cite{Gaido2020,zhang-etal-2021-beyond} might further improve the translation performance.

\subsection{Automatic segmentation repairs improper segmentation in ST corpus segments}

\begin{figure*}[t]
\centering
\includegraphics[keepaspectratio, width=16cm]{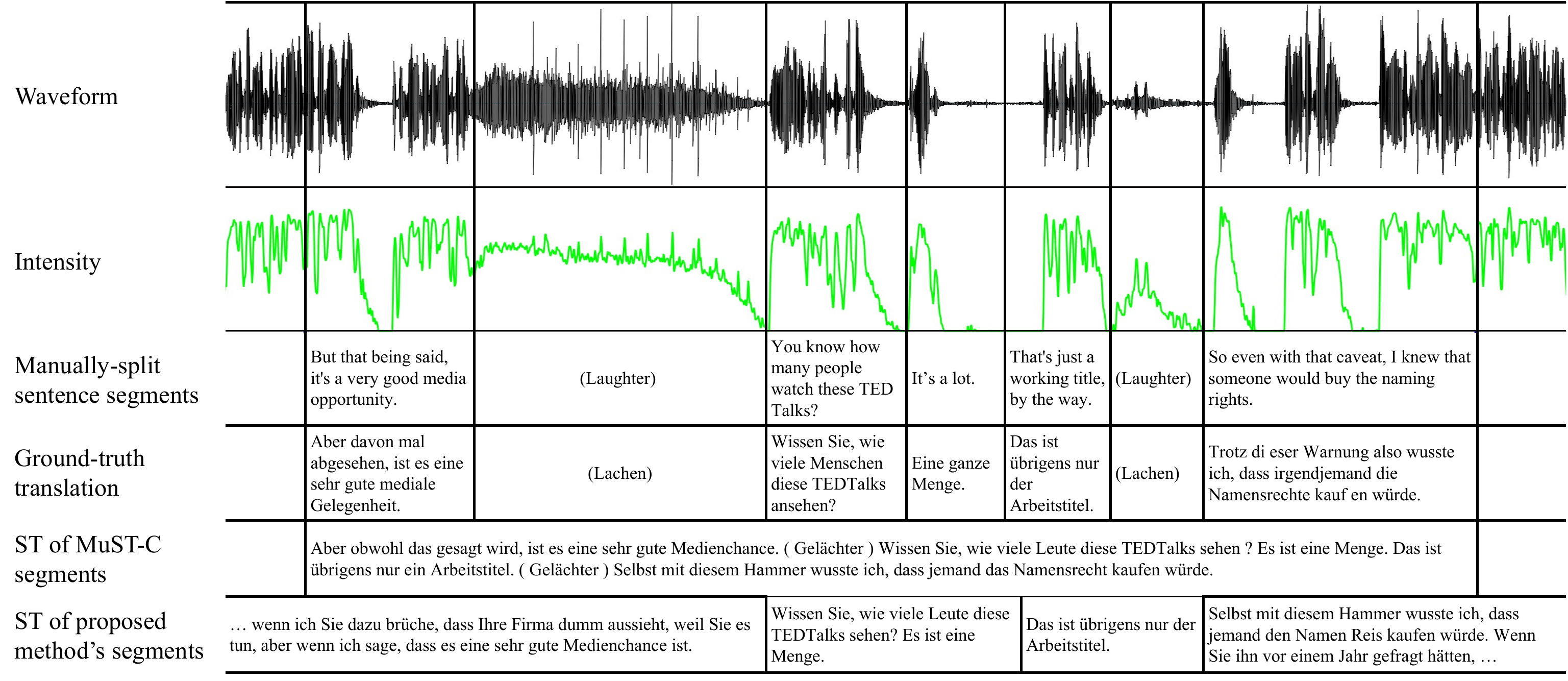}
\caption{Examples of segmentation and its ST result: Vertical lines indicate the split position of a waveform.}
%  Third row from the bottom shows manually-split sentence segments.
\label{fig:case_study}
\end{figure*}

Although the prediction accuracy of the best SFC model has room for improvement (Table \ref{tab:sfc_acc}), it maintains a BLEU score as high as 97\% of the sentence-aligned speech segmentation.
We conducted a case study, hypothesizing that automatic segmentation can repair the incorrect segmentation in the ST corpus.
Fig. \ref{fig:case_study} shows examples of MuST-C and automatic segmentation and their ST results.
Compared to the sentence segments, it can be seen that MuST-C segmentation often overlooked sentence boundaries.
This was caused by audience laughter and short pauses between utterances.
Such under-segmentation increases segment lengths, which often degrade the translation accuracy.
In contrast, the proposed method predicted the sentence segment boundaries more accurately than MuST-C segmentation.
Of course, there are over- and under-segmentations by the proposed method.
However, in some cases, the proposed method outperformed MuST-C segmentation, leading to competitive translation results.

% oracleに近い分布
% contextについてもここで言及
% segmentが長めになりやすいpDACとかpSTRMは、contextである程度BLEUを稼いでる、と仮定できる
% pTHRはcontext-awareなSTと組み合わせて更に向上が見込める
% （将来）

% \subsection{Applicability}
% \subsubsection*{\bf Cross-domain}
% \subsubsection{\bf Distant language}
% \subsubsection{\bf Multi-linguality}
% \subsection{Relationship Between Segmentation Accuracy and Translation Performance}
% \subsection{Qualitative Analysis}

% Other language pair: とりあえずここに置いてみる
\begin{table*}[]
\caption{Results in BLEU by $SHAS$ and $SHAS+FTPT$ for the 8 language pairs of MuST-C v1 tst-COMMON. Four numbers separated by slashes are the results of different algorithms (pDAC/pSTRM/pTHR/pTHR+MA).}
\label{tab:multilingual_results}
\centering
\begin{tabular}{|l||c|c|c|c|}
\hline
          & MuST-C v1 En-De                   & MuST-C v1 En-Es                   & MuST-C v1 En-Fr                   & MuST-C v1 En-It                   \\ \hline
Sentence-aligned    & 23.31                   & 28.12                   & 33.93                   & 24.20                   \\ \hline
SHAS (\textit{middle}) & 21.75/21.42/19.69/21.93 & 26.69/26.43/22.84/25.73 & 31.36/30.84/28.69/31.70 & 22.67/22.22/19.68/21.93 \\ 
SHAS+FTPT (\textit{large+all}) & 22.36/21.85/22.61/\textbf{22.65} & 26.78/26.68/26.75/\textbf{26.86} & 31.96/31.50/32.03/\textbf{32.07} & \textbf{23.04}/22.61/22.79/22.95 \\ \hline \hline
          & MuST-C v1 En-Nl                   & MuST-C v1 En-Pt                   & MuST-C v1 En-Ro                   & MuST-C v1 En-Ru                   \\ \hline
Sentence-aligned    & 27.92                   & 30.38                   & 21.93                   & 15.57                   \\ \hline
SHAS (\textit{middle}) & 26.28/25.97/22.97/25.20 & 28.55/28.30/25.85/28.39 & 20.31/19.96/18.62/20.38 & 14.59/14.35/12.57/14.19 \\
SHAS+FTPT (\textit{large+all}) & \textbf{26.74}/26.44/26.48/26.20 & 29.15/28.74/29.02/\textbf{29.28} & \textbf{21.05}/20.50/20.91/20.86 & \textbf{14.61}/14.31/14.43/14.53 \\ \hline
\end{tabular}
\end{table*}

% Other domain: とりあえずここに置いてみる
\begin{table}[!t]
\caption{Results in BLEU by $SHAS$ and $SHAS+FTPT$ for En-De of Europarl-ST. Four numbers separated by slashes are the results of different algorithms (pDAC/pSTRM/pTHR/pTHR+MA).}
\label{tab:other_domain_results}
\centering
\begin{tabular}{|l|c|}
\hline
 & Europarl-ST En-De \\ \hline \hline
Sentence-aligned  & 26.17 \\ \hline
SHAS (\textit{middle}) & 24.35/22.77/24.06/24.65 \\ \hline
SHAS+FTPT (\textit{large+all}) & 25.13/24.15/\textbf{25.68}/25.48 \\ \hline
\end{tabular}
\end{table}

\subsection{Testing on other datasets}
Table \ref{tab:multilingual_results} shows the results of applying a combination of two SFC models (baseline SHAS and the proposed SHAS+FTPT) and four algorithms (pDAC, pSTRM, pTHR, and pTHR+MA) to the 8 languages of MuST-C v1.
No hyperparameters were fine-tuned in these tests, and all segmentation methods were applied with exactly the same configuration as used in his MuST-C v2 en-de.
% For these tests, we did not fine-tune any hyperparameters and applied all segmentation methods with the exact same settings as used for MuST-C v2 en-de. 
From the model perspective, SHAS+FTPT consistently outperformed SHAS. 
Observations from the algorithm side were also in line with our main experiments: the translation accuracy of the SHAS model was particularly low when using pTHR, while pTHR+MA showed translation accuracy comparable to pDAC. These results lead us to conclude that our proposed method is effective for different target languages.
Table \ref{tab:other_domain_results} shows the test results for a different domain, Europarl-ST.
The improvement in translation accuracy when using pTHR (from 24.06 to 25.68) suggests that our approach of unfreezing wav2vec brings about an enhancement in generalization ability, rather than mere overfitting to the task.

\section{Conclusion}
In this study, we addressed a problem caused by the state-of-the-art speech segmentation method, SHAS, which tends to generate overly long segments, degrading the quality and time efficiency of speech translation (ST).
We extended SHAS to improve ST translation accuracy and efficiency by splitting speech into shorter segments that correspond to sentences.
We introduced a simple segmentation algorithm using the moving average of SFC predictions without relying on length heuristics.
We also introduced efficient fine-tuning of wav2vec 2.0 to improve the SFC of SHAS and investigated the effects of model size and trainable parameters on prediction performance.
% Experiments using ST corpora showed that each technique effectively improves the translation accuracy, and that combining them yields the best accuracy improvement.
% Our proposed method not only improved the translation accuracy; it also improved ST's time efficiency by generating shorter segments.
Experiments using ST corpora showed that the proposed segmentation algorithm improves ST's time efficiency by generating shorter segments while maintaining translation quality comparable to existing algorithms.
Experimental results also showed that fine-tuning wav2vec 2.0 improves the accuracy of SFC, with a concomitant significant improvement in ST quality.
\par
Future research will focus on the proposed method for simultaneous speech translation.
It will also investigate adapting to the noisy and multi-speaker environments, optimizing segmentation to maximize translation accuracy, and combining with context-aware ST.

\section*{Acknowledgments}
Part of this work was supported by JST SPRING Grant Number JPMJSP2140 and JSPS KAKENHI Grant Numbers JP21H05054 and JP21H03500.

% \section{References Section}
% You can use a bibliography generated by BibTeX as a .bbl file.
%  BibTeX documentation can be easily obtained at:
%  http://mirror.ctan.org/biblio/bibtex/contrib/doc/
%  The IEEEtran BibTeX style support page is:
%  http://www.michaelshell.org/tex/ieeetran/bibtex/

\bibliographystyle{IEEEtran}
\bibliography{IEEE_TASLP_ryo_fukuda}

\newpage
\appendix

\subsection{Hyperparameters}
\label{app:best_parameters}

% \begin{table}[h]
% \centering
% \caption{Best parameters of pDAC chosen by dev set and results in BLEU and number of segments (\#seg).}
% \label{tab:param_pdac}
% \begin{tabular}{|l|c|c|c|}
% \hline
% Model          & BLEU  & \# seg & max \\ \hline\hline
% lna\_l16\_ft0  & 25.42 & 1073   & 26            \\
% lna\_l16\_ft4  & 25.84 & 1656   & 16            \\
% lna\_l16\_ft8  & 25.75 & 1268   & 22            \\
% lna\_l16\_ft16 & 25.73 & 1855   & 14            \\ \hline
% lna\_l24\_ft0  & 24.41 & 1031   & 28            \\
% lna\_l24\_ft6  & 25.73 & 1636   & 16            \\
% lna\_l24\_ft12 & 25.89 & 1389   & 20            \\
% lna\_l24\_ft24 & 25.95 & 2279   & 10            \\ \hline
% \end{tabular}
% \end{table}

\begin{table}[h]
\centering
\caption{Best parameters of each algorithm chosen by dev set and results in BLEU and number of segments (\#seg).}
\label{tab:combined}

\begin{subtable}{0.48\textwidth}
\centering
\caption{pDAC}
\label{tab:param_pdac}
\begin{tabular}{|l|c|c|c|}
\hline
Model          & BLEU  & $\# seg$ & $max$ \\ \hline\hline
lna\_l16\_ft0  & 25.42 & 1073   & 26            \\
lna\_l16\_ft4  & 25.84 & 1656   & 16            \\
lna\_l16\_ft8  & 25.75 & 1268   & 22            \\
lna\_l16\_ft16 & 25.73 & 1855   & 14            \\ \hline
lna\_l24\_ft0  & 24.41 & 1031   & 28            \\
lna\_l24\_ft6  & 25.73 & 1636   & 16            \\
lna\_l24\_ft12 & 25.89 & 1389   & 20            \\
lna\_l24\_ft24 & 25.95 & 2279   & 10            \\ \hline
\end{tabular}
\end{subtable}

\vspace{1cm}

\begin{subtable}{0.48\textwidth}
\centering
\caption{pSTRM}
\label{tab:param_pstrm}
\begin{tabular}{|l|c|c|c|}
\hline
Model          & BLEU   & $\# seg$ & $max$ \\ \hline\hline
lna\_l16\_ft0  & 25.11 & 953    & 28  \\
lna\_l16\_ft4  & 25.57  & 1186   & 22  \\
lna\_l16\_ft8  & 25.52 & 1294   & 20  \\
lna\_l16\_ft16 & 25.71 & 1300   & 20  \\ \hline
lna\_l24\_ft0  & 25.18 & 1648   & 16  \\
lna\_l24\_ft6  & 25.7 & 1584   & 16  \\
lna\_l24\_ft12 & 25.58  & 1304   & 20  \\
lna\_l24\_ft24 & 25.70   & 1292   & 20  \\ \hline
\end{tabular}
\end{subtable}

\vspace{1cm}

\begin{subtable}{0.48\textwidth}
\centering
\caption{pTHR}
\label{tab:param_pthr}
\begin{tabular}{|l|c|c|c|c|}
\hline
Model          & BLEU  & $\# seg$ & $thr$ & $ma$ \\ \hline
lna\_l16\_ft0  & 23.54 & 3639   & 0.1 & 0  \\
lna\_l16\_ft4  & 25.67 & 2353   & 0.1 & 0  \\
lna\_l16\_ft8  & 25.92 & 2703   & 0.2 & 0  \\
lna\_l16\_ft16 & 26.13 & 2525   & 0.2 & 0  \\ \hline
lna\_l24\_ft0  & 21.15 & 4860   & 0.1 & 0  \\
lna\_l24\_ft6  & 25.73 & 2588   & 0.2 & 0  \\
lna\_l24\_ft12 & 26.26 & 2638   & 0.2 & 0  \\
lna\_l24\_ft24 & 26.28 & 2149   & 0.1 & 0  \\ \hline
\end{tabular}
\end{subtable}

\vspace{1cm}

\begin{subtable}{0.48\textwidth}
\centering
\caption{pTHR+MA}
\label{tab:param_pthr_ma}
\begin{tabular}{|l|c|c|c|c|}
\hline
Model          & BLEU  & $\# seg$ & $thr$ & $ma$  \\ \hline
lna\_l16\_ft0  & 25.73 & 1944   & 0.1 & 0.2 \\
lna\_l16\_ft4  & 25.96 & 2055   & 0.1 & 0.1 \\
lna\_l16\_ft8  & 26.17 & 2464   & 0.2 & 0.1 \\
lna\_l16\_ft16 & 26.27 & 1981   & 0.1 & 0.1 \\ \hline
lna\_l24\_ft0  & 24.78 & 1816   & 0.1 & 0.4 \\
lna\_l24\_ft6  & 26.13 & 1914   & 0.1 & 0.1 \\
lna\_l24\_ft12 & 26.15 & 2005   & 0.1 & 0.1 \\
lna\_l24\_ft24 & 26.30 & 2044   & 0.1 & 0.1 \\ \hline
\end{tabular}
\end{subtable}

\end{table}

\vfill

\end{document}